\documentclass[superscriptaddress,twocolumn,showpacs,amsmath,amssymb,nofootinbib]{revtex4}
\usepackage{graphicx}
\usepackage{dcolumn}
\usepackage{bm}

\begin{document}

\title{Mass dispersion in transfer reactions with a stochastic mean-field theory}

\author{Kouhei Washiyama}
\affiliation{GANIL, Bd Henri Becquerel, BP 55027, 14076 Caen Cedex 5, France}

\author{Sakir Ayik}
\affiliation{Physics Department, Tennessee Technological University,
Cookeville, Tennessee 38505, USA}
\affiliation{Physics Department,
Middle East Technical University, 06531 Ankara, Turkey}

\author{Denis Lacroix}
\affiliation{GANIL, Bd Henri Becquerel, BP 55027, 14076 Caen Cedex 5, France}

\date{\today}

\begin{abstract}
Nucleon transfer in symmetric heavy-ion reactions at energies below the Coulomb barrier is
investigated in the framework of a microscopic stochastic mean-field theory.
While mean-field alone is known to significantly underpredict the dispersion of the fragment mass
distribution, a considerable enhancement of the dispersion is obtained in the stochastic mean-field theory.
The variance of the fragment mass distribution deduced from the stochastic theory scales with the number
of exchanged nucleon. Therefore, the new approach provides the first fully microscopic theory  
consistent with the phenomenological analysis of the experimental data. 
\end{abstract}


\maketitle


The mean-field theory, otherwise known as the time-dependent Hartree-Fock (TDHF) model~\cite{koonin80,
negele82, goeke, davis, Sim08,davies78}, has been widely used in describing reaction dynamics at low energies
in nuclear collisions and other many-body systems. It is well known that the mean-field theory provides
a good description of mean values of one-body observables in low energy
reactions. However, it completely fails in the description of the dynamics of fluctuations of one-body observables. 
During the past decades, large efforts have been devoted to overcome this difficulty and to develop transport
theories that are able to describe not only mean values but also fluctuations 
(for a review see \cite{Abe96,Lac04}). Among them, the variational principle proposed by Balian and V\'en\'eroni (BV)
appears as one of the most promising methods~\cite{Bal84,balian84,Mar85}. However, even nowadays it remains difficult to
apply \cite{broomfield08,Ste09}. More than 30 years after the first application of TDHF, 
the absence of practical solution to include fluctuations beyond mean-field in a fully microscopic 
framework strongly restricts applications of mean-field based theories. 

There are mainly two mechanisms for density fluctuations: (i) collisional fluctuations generated 
by two-body collisions and (ii) one-body mechanism or mean-field fluctuations. 
At low bombarding energies, the mean-field fluctuations provide the dominant mechanism for
fluctuations of collective motion and collisional effects could be neglected. 
Restricting the treatment at low energies, recently, a stochastic
mean-field (SMF) approach has been proposed for nuclear dynamics~\cite{Ayik2}. It was demonstrated
that the approach
incorporates the one-body dissipation and associated fluctuation mechanism in accordance with the quantal
dissipation-fluctuation relation. Furthermore, in the limit of small amplitude fluctuations, the SMF
approach gives the same result for dispersion of one-body observables as that of
the formula derived from the BV approach~\cite{Bal84,balian84}.
Therefore, the SMF approach provides a powerful
tool for describing low energy nuclear processes including induced fission, heavy-ion fusion near barrier
energies, deep-inelastic collisions, and spinodal decomposition of nuclear matter~\cite{Ayik08}. 
In this work, by extending the previous work~\cite{ayik09}, we
study nucleon exchange in low-energy nuclear
collisions and calculate the dispersion of the fragment mass distribution~\cite{negele82,goeke,davis,davies78,dasso79}.
Diffusion coefficients for nucleon exchange as well as for momentum transfer extracted from the SMF approach have
the same structure as the result familiar from the phenomenological nucleon exchange model.

The SMF approach is based on a very appealing stochastic model proposed for
describing deep-inelastic heavy-ion collisions and sub-barrier
fusion ~\cite{Esb78, Dasso1,Dasso2}. In that model, dynamics of
relative motion is coupled to collective surface modes of colliding
ions and treated in a classical framework. The initial quantum zero-point 
fluctuations are incorporated into the calculations in a
stochastic manner by generating an ensemble of events according to
the initial distribution of collective modes. In the mean-field
evolution, couplings of relative motion with all other collective and
non-collective modes are automatically taken into account. In the
stochastic extension of the mean-field approach, the zero-point (and
thermal) fluctuations of the initial state are taken into account in
a stochastic manner, which is similar to the spirit presented in Refs.~\cite{Esb78,Dasso1,Dasso2}. 
The initial fluctuations are simulated by considering an ensemble of
initial single-particle density matrices. In this manner, the single
Slater determinantal description is replaced by a superposition of
Slater determinants. A member of the ensemble, 
indicated by event label $\lambda$, can be expressed as
\begin{eqnarray}
\label{eq:density}
\rho^\lambda(\boldsymbol{r},{\boldsymbol{r}}^{\prime},t) = \sum\limits_{ij \sigma\tau}
\Phi_{i \sigma\tau}^\ast(\boldsymbol{r},t;\lambda )\rho_{ij}^\lambda(\sigma\tau)
\Phi_{j \sigma\tau}(\boldsymbol{r}^{\prime},t;\lambda ),
\end{eqnarray}
where summations $i$ and $j$ run over a complete set of single-particle states
$\Phi_{i \sigma\tau}(\boldsymbol{r},t;\lambda )$, and  $\sigma$ and $\tau$ denote
spin and isospin quantum numbers. According to the description of the SMF approach
~\cite{Ayik2}, the element of density matrix, $\rho_{ij}^\lambda(\sigma\tau)$
are assumed to be time-independent random Gaussian numbers with mean value
$\overline{\rho_{ij}^\lambda}(\sigma\tau)=\delta_{ij}n_i^{\sigma\tau}$
and the variance of the fluctuating part $\delta\rho_{ij}^\lambda(\sigma\tau)$
specified by
\begin{multline}
\overline{\delta\rho_{ij}^\lambda(\sigma\tau)
\delta\rho_{j'i'}^\lambda( {\sigma}'{\tau}')} \\
=\frac{1}{2}\delta_{j{j}'}\delta_{i{i}'}\delta_{\tau {\tau}'}\delta_{\sigma {\sigma}'}
\left[n_i^{\sigma\tau}(1 - n_j^{\sigma\tau}) + n_j^{\sigma\tau}(1 - n_i^{\sigma\tau})\right].
\label{variance}
\end{multline}
Here, $n_i^{\sigma\tau}$ denotes the average single-particle occupation factor. At zero temperature
occupation factors are $0$ and $1$, and at finite temperature they are determined by the Fermi-Dirac distribution.
The great advantage of the SMF theory is that each Slater determinant $\lambda$
evolves independently from each other following the time evolution
of its single-particle wave-functions in its self-consistent mean-field Hamiltonian, denoted by $h(\rho^\lambda)$,
according to
\begin{eqnarray}
\label{eq:spwf}
i\hbar \frac{\partial }{\partial t}\Phi_{i \sigma\tau} (\boldsymbol{r},t;\lambda )
= h(\rho^\lambda )\Phi_{i \sigma\tau}(\boldsymbol{r},t;\lambda ).
\end{eqnarray}

Following Refs.~\cite{Washiyama1,Washiyama2}, we project the mean-field evolution
on a collective degree of freedom associated with nucleon transfer.
For the projection, it is useful to introduce the Wigner distribution for the event $\lambda$
defined as a partial Fourier transform of the density matrix,
\begin{multline}
f^\lambda(\boldsymbol{r},\boldsymbol{p},t)= \int d^3s\exp\left( - \frac{i}{\hbar}
\boldsymbol{p}\cdot\boldsymbol{s}\right)  \\
\times\sum_{ij \sigma\tau}
\Phi_{j \sigma\tau}^\ast\left(\boldsymbol{r} + \frac{\boldsymbol{s}}{2},t;\lambda\right)
\rho_{ji}^\lambda(\sigma\tau)
\Phi_{i \sigma\tau}\left(\boldsymbol{r} - \frac{\boldsymbol{s}}{2},t;\lambda\right).
\label{eq:wigner}
\end{multline}
In this work, we focus on the particular case of head-on collisions along the $x$-axis.
We indicate the position of the separation plane between the two collision partners at $x=x_0$.
Then, the mass number of the projectile-like fragment in the event $\lambda$ is defined by
\begin{eqnarray}
\label{eq:mass}
A_P^\lambda(t)=\int\frac{d^3rd^3p}{(2\pi \hbar)^3}\theta(x-x_0)
f^\lambda(\boldsymbol{r},\boldsymbol{p} ,t).
\end{eqnarray}
Other macroscopic variables such as the separation distance between fragments and the associated momentum
can be defined in a similar manner (see \cite{Washiyama2}).
In the diffusion model, time evolution of the mass number
of the projectile-like fragment $A_P^\lambda $ is described by a Langevin equation~\cite{Randrup2},
\begin{eqnarray}
\label{eq:langevin}
\frac{d}{dt}A_P^\lambda = v(A_P^\lambda, t) + \xi_A^\lambda(t),
\end{eqnarray}
where $v(A_P^\lambda ,t)$ denotes the drift coefficient
for nucleon transfer.
Ignoring memory effects, we consider the quantity $\xi_A^\lambda (t)$ as a
Gaussian white noise, which is determined with zero mean value $\overline{\xi_A^\lambda}(t)=0$
and a correlation function,
\begin{eqnarray}
\label{eq:correlationmass}
\overline{\xi_A^\lambda (t)\xi_A^\lambda ({t}')} = 2\delta(t -{t}')D_{AA},
\end{eqnarray}
where $D_{AA}$ is the diffusion coefficient associated with nucleon exchange.
In order to extract the diffusion coefficient, we calculate the rate of
change of $A_P^\lambda $ employing the SMF equations.
The rate of change of  $A_P^\lambda$ involves only the kinetic part of the mean-field Hamiltonian and it can
be expressed in terms of the reduced Wigner distribution on the window as
\begin{eqnarray}
\frac{d}{dt}A_{P}^{\lambda}
= - \int \frac{dp_x}{2\pi\hbar} \frac{p_x}{m} f^{\lambda}(x,p_x ,t)\bigr\vert_{x = x_{0}},
\end{eqnarray}
where the reduced Wigner distribution $f^{\lambda} (x,p_x ,t)$ is obtained by
integrating over the phase-space variables $y$, $z$, $p_y$, and $p_z$ according to
\begin{eqnarray}
\label{eq:wignerx}
f^{\lambda}(x,p_x ,t) = \iint dydz\frac{dp_y dp_z}{(2\pi\hbar)^2}
f^{\lambda}(\boldsymbol{r},\boldsymbol{p} ,t).
\end{eqnarray}
Small fluctuations of the mass number are connected to small amplitude
fluctuations of the Wigner distribution according to
\begin{eqnarray}
\label{eq28}
\frac{d}{dt}\delta A_{+}^{\lambda} =
-\int\frac{dp_x}{2\pi\hbar}\frac{p_x}{m}\delta f^{\lambda}(x,p_x ,t)\bigr\vert_{x = x_{0}}
= \xi_A^{\lambda}(t).
\end{eqnarray}
In Ref.~\cite{ayik09}, we derived an expression for the correlation function of the reduced
Wigner distribution in the semi-classical approximation. Employing the result derived in that
reference, we have
the following expression for the nucleon diffusion coefficient,
\begin{multline}
D_{AA}(t) = \int\frac{dp_x }{2\pi \hbar}
\frac{\vert p_x \vert}{m}\frac{1}{2} \\
\times\sum_{\sigma\tau} \bigl\{f_P^{\sigma\tau}(x_0,p_x ,t)
\left[1 - f_T^{\sigma\tau}(x_0,p_x,t)\right/\Omega(x_0,t)] \\
+ f_T^{\sigma\tau}(x_0,p_x,t)\left[1 - f_P^{\sigma\tau}(x_0,p_x,t)\right]/\Omega(x_0,t)\bigr\}.
\label{eq:dAA}\end{multline}
Here $\Omega(x_0,t)$ is the phase-space volume over the window and
\begin{multline}
 f_{P/T}^{\sigma\tau}(x_0,p_x ,t) = \iint dydz
\int ds_x\exp\left( - \frac{i}{\hbar}p_x s_x\right) \\
\times \sum_{i\in P/T}
\Phi_{i \sigma\tau}^\ast \left(x + \frac{s_x}{2},y,z,t\right)
n_{i}^{\sigma\tau} \Phi_{i \sigma\tau}\left(x - \frac{s_x}{2},y,z,t\right)
\label{eq:wignerPT}
\end{multline}
is the averaged value of the reduced Wigner distribution
associated with single-particle wave functions originating from
the projectile/target. Details on the determination of $\Omega(x_0,t)$
can be found in Ref.~\cite{ayik09}.

We note that the expression of the diffusion coefficient has the same form
as given by the phenomenological nucleon exchange model in Ref.~\cite{Feldmeier}.
We also note that diffusion coefficients not only for nucleon exchange but also
associated with other macroscopic variables are evaluated in terms of the
average evolution specified by the standard TDHF evolution. In computations,
to employ fully quantum mechanical expression for the reduced Wigner distribution
does not provide a consistent description
since the diffusion coefficient is derived in the semi-classical approximation.
A semi-classical form of the reduced Wigner
distribution can be obtained by approximating the $s_{x}$ dependence of the integrand
of the expression (\ref{eq:wigner}) by a Gaussian. The mean value and the second moment of this Gaussian
are determined by carrying out  a Taylor expansion of the integrand up to second
order in $s_{x}$.

We carry out calculations for head-on collisions of symmetric
$^{40}$Ca+$^{40}$Ca, $^{56}$Ni+$^{56}$Ni and $^{90}$Zr+$^{90}$Zr
systems at energies just below the Coulomb barrier.
Calculations are performed using the three-dimensional TDHF code
developed by P. Bonche and co-workers with the SLy4d Skyrme
effective force~\cite{kim97} (technical details are given in
Ref.~\cite{Washiyama1}). Colliding ions approach each other,
exchange a number of nucleons, and then re-separate. In symmetric
collisions by TDHF, there is no net nucleon transfer, i.e., drift is zero.
According to the Langevin equation, the variance $
\sigma_{AA}^{2}(t)=\overline{\delta A_{P}^{\lambda}\delta
A_{P}^{\lambda}}$ of fragment mass distribution, neglecting
contributions from the drift term, is related to the diffusion
coefficient according to \cite{Randrup2,Randrup82}
\begin{eqnarray}
\sigma^2_{AA}(t) \simeq 2 \int_0^t D_{AA}(s)ds=N_{\rm exc}(t).
\label{eq:sigma}
\end{eqnarray}
In this expression $N_{\rm exc}(t)$ denotes the accumulated total number of exchanged nucleons
until time $t$. The relation $ \sigma_{AA}^{2}(t)=N_{\rm exc}(t)$ follows from nucleon exchange
model and it was often used to analyze the experimental data. In the SMF model, we
can calculate the both sides of this relation independently.
\begin{figure}
\includegraphics[width=0.64\linewidth, clip]{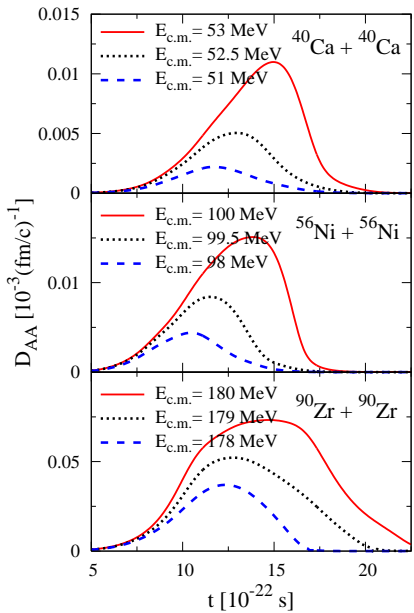}
\caption{(Color online) Evolution of  diffusion coefficient calculated in SMF approach for
$^{40}$Ca${}+^{40}$Ca (top), $^{56}$Ni${}+^{56}$Ni (middle) and
$^{90}$Zr${}+^{90}$Zr (bottom) collisions at different center-of-mass energies.
}
\label{fig:DAAtime}
\end{figure}

Figure~\ref{fig:DAAtime} illustrates the dependence of diffusion coefficients for collision of three different symmetric
systems at different center-of-mass energies. The Coulomb barrier energies, which are obtained 
in the frozen density approximation~\cite{denisov02},
are 54.7 MeV, 103 MeV, and 184 MeV for the $^{40}$Ca, $^{56}$Ni, and $^{90}$Zr systems, respectively.
The magnitude of diffusion coefficient essentially depends on the size of the window, the larger the
window the larger the rate of change of nucleon exchange \cite{Randrup82}. At a given center-of-mass energy, 
diffusion coefficient becomes maximum at the turning point where the size of window is the largest.  
Also, as seen from the figure, due to increasing overlap of the projectile and target, the magnitude of 
the diffusion coefficient increases with energy.  
\begin{figure}
\includegraphics[width=0.64\linewidth, clip]{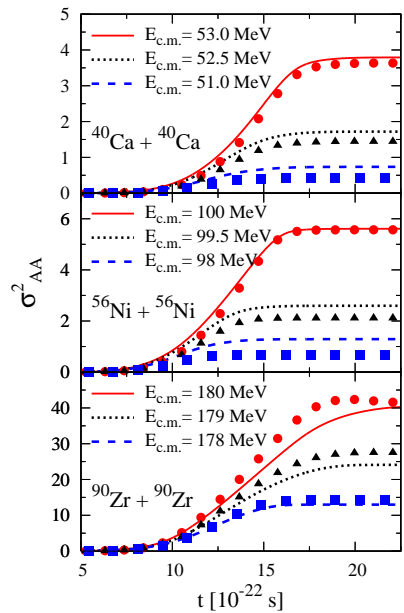}
\caption{(Color online) Evolution of $\sigma_{AA}^2$ calculated in SMF approach for
 $^{40}$Ca${}+^{40}$Ca (top), $^{56}$Ni${}+^{56}$Ni (middle) and
$^{90}$Zr${}+^{90}$Zr (bottom) collisions at different center-of-mass energies.
Number of exchanged nucleons is superimposed
by the filled-circles, filled-squares, and filled-triangles
from high to low energies.}
\label{fig:SIGMAtime}
\end{figure}

Figure~\ref{fig:SIGMAtime} illustrates the variances of the fragment mass distributions as a function
of time for the same symmetric systems at the same center-of-mass energies 
as those in Fig.~\ref{fig:DAAtime}. Lines are the results obtained 
by integration of diffusion coefficient in Eq.~(\ref{eq:sigma}). In each case, the corresponding evolution 
of the number of exchanged nucleons is superimposed
by the filled-circles, filled-squares, and filled-triangles from high to low energies.
We also calculate the variances of fragment mass distribution using standard TDHF approach. 
In Table~\ref{table:deltaA}, for the same systems at the same energies, the asymptotic values 
$\sigma^2_{\rm TDHF}(+\infty)$ of the variances 
obtained from TDHF is compared with the asymptotic values $\sigma^2_{AA}(+\infty)$ calculated in the SMF approach.
In the same table, the asymptotic values of the number of exchanged nucleons $N_{\rm exc} (+\infty)$ are also given.

\begin{table}[htbp]
\begin{center}
\caption{Asymptotic values of the fragment mass variances for
$^{40}$Ca${}+^{40}$Ca, $^{56}$Ni${}+^{56}$Ni, $^{90}$Zr${}+^{90}$Zr collisions 
in SMF ($\sigma^2_{AA}$) and TDHF ($\sigma^2_{\rm TDHF}$). Asymptotic values of the number
of exchanged nucleons are also given in the last column.
}
\label{table:deltaA}
\begin{tabular}{cc|c c c}
\hline
Reaction & $E_{\rm c.m.}$[MeV] & $\sigma^2_{\rm TDHF}(+\infty)$  & $\sigma^2_{\rm AA}(+\infty)$  & $N_{\rm exc} (+\infty)$  \\
\hline 
\hline
$^{40}$Ca+$^{40}$Ca & 51.0 & 0.004  & 0.730 & 0.432 \\
                    & 52.5 & 0.008  & 1.718 & 1.441\\
                    & 53.0 & 0.008  & 3.790 & 3.634\\
\hline
$^{56}$Ni+$^{56}$Ni & 98  & 0.024  & 1.288 & 0.667\\
                    & 99.5& 0.088  & 2.594 & 2.111\\
                    & 100 & 0.142  & 5.606 & 5.567\\
\hline
$^{90}$Zr+$^{90}$Zr & 178 & 0.774 & 12.98  & 14.19\\
                    & 179 & 1.046 & 24.09  & 27.51\\
                    & 180 & 1.085 & 40.40  & 41.56\\
\hline
\end{tabular}
\end{center}
\end{table}

The semi-empirical relation $ \sigma_{AA}^{2}(t) \simeq N_{exc}(t)$ has been extensively used 
to analyze experimental data \cite{Fre84,Ada94}. 
As seen in Table~\ref{table:deltaA}, the mass variance estimated from SMF is consistent 
with this relation.
Looking at the asymptotic values, in all cases, 
the TDHF results $\sigma^2_{\rm TDHF}(+\infty)$ for the variance are much smaller than
the results $\sigma^2_{AA}(+\infty)$ obtained in the SMF approach and also $N_{\rm exc} (+\infty)$.
The failure of TDHF on the description of variances of the fragment mass distribution
has been recognized for a long time as a major limitation of the mean-field theory. 
It appears that the SMF approach cures this shortcoming
of the mean-field theory. As seen from Fig.~\ref{fig:SIGMAtime}, not only the asymptotic value of $\sigma^2_{AA}$ but also the entire
time evolution is very close to the evolution of $N_{\rm exc}(t)$. The small differences at energies much below the 
barrier may arise from the estimation of the phase-space volume on the window and/or from the Gaussian
approximation for the reduced Wigner distribution.

In summary, we investigate the variances of fragment mass distributions in 
heavy-ion collisions at energies near the Coulomb barrier employing the microscopic SMF approach. 
By projecting the SMF equation on the mass-asymmetry macroscopic variable, we deduce
the diffusion coefficient associated with nucleon exchange. The expression of the diffusion coefficient
has a similar structure with those familiar from the phenomenological nucleon exchange model. Comparison between
the calculated variance and the number of exchanged nucleon supports a strong confirmation for the 
fact that the SMF approach provides a realistic description of dissipation and fluctuation dynamics
at low energies. The stochastic extension of the mean-field theory provides a practical solution to the 
estimate of fluctuations of observables at low energies.  It can be applied not only to nuclear
dynamics but also for the description of fluctuating dynamics of many-body problems in other areas
of physics. 


D.L. and K.W. gratefully acknowledge TUBITAK for financial support and METU
for warm hospitality extended to them during their visit. S.A. gratefully
acknowledges CNRS and TUBITAK for financial support and
GANIL for warm hospitality extended to him during his visit.  This work is
supported in part by US DOE Grant DE-FG05-89ER40530.


\begin{thebibliography}{99}

\bibitem{koonin80} S. E. Koonin, Prog. Part. Nucl. Phys. {\bf 4}, 283 (1980).

\bibitem{negele82} J. W. Negele, Rev. Mod. Phys. {\bf 54}, 913 (1982).

\bibitem{goeke} K. Goeke and P.-G. Reinhard, Time-Dependent Hartree-Fock and Beyond, Bad
Honnef, Germany, 1982.

\bibitem{davis} K. T. D. Davis, K. R. S. Devi, S. E. Koonin, and M. Strayer, Treatise in
Heavy-Ion Science, ed. D. A. Bromley, Nuclear Science V-4, Plenum, New York, 1984

\bibitem{Sim08} C. Simenel, B. Avez, and D. Lacroix, Lecture notes of the "International Joliot-Curie School",
Maubuisson, September 17-22, 2007, arXiv:0806.2714.

\bibitem{davies78} K. T. R. Davies, V. Maruhn-Rezwani, S. E. Koonin, and J. W. Negele,
Phys. Rev. Lett. {\bf 41}, 632 (1978).

\bibitem{Abe96}
Y. Abe, S. Ayik, P.-G. Reinhard, and E. Suraud, Phys. Rep. {\bf 275}, 49 (1996).

\bibitem{Lac04}  D. Lacroix, S. Ayik, and Ph. Chomaz,
Prog. Part. Nucl. Phys. {\bf 52}, 497 (2004).


\bibitem{Bal84} R. Balian and M. V\'en\'eroni, Phys. Lett. {\bf B136}, 301 (1984).

\bibitem{balian84} R. Balian, P. Bonche, H. Flocard, and M. V\'en\'eroni,
Nucl. Phys. {\bf A428}, 79c (1984).

\bibitem{Mar85} J. B. Marston and S. E. Koonin, Phys. Rev. Lett. {\bf 54}, 1139 (1985).

\bibitem{broomfield08} J. M. A. Broomfield and P. D. Stevenson,
J. Phys. G {\bf 35}, 095102 (2008).

\bibitem{Ste09} P. D. Stevenson and J. M. A. Broomfield, proceedings of XXXII Symposium 
on Nuclear Physics, Cocoyoc, Mexico (2009), arXiv:0903.0130.

\bibitem{Ayik2} S. Ayik, Phys. Lett. {\bf B658}, 174 (2008).

\bibitem{Ayik08} S. Ayik, N. Er, O. Yilmaz, and A. Gokalp, Nucl. Phys. {\bf A 812}(2008) 44.

\bibitem{ayik09} S. Ayik, K. Washiyama, and D. Lacroix, Phys. Rev. C {\bf 79},
054606, (2009).

\bibitem{dasso79} C. H. Dasso, T. D{\o}ssing, and H. C. Pauli,
Z. Phys. A {\bf 289}, 395 (1979).

\bibitem{Esb78} H. Esbensen, A. Winther, R. A. Broglia, and C. H. Dasso, Phys. Rev. Lett. {\bf 41}, 296 (1978).

\bibitem{Dasso1} C. H. Dasso, Proc. Second La Rapida Summer School on Nuclear Physics,
eds. M. Lozano and G. Madurga, World Scientific, Singapore, 1985.

\bibitem{Dasso2} C. H. Dasso and R. Donangelo, Phys. Lett. {\bf B276}, 1 (1992).

\bibitem{Washiyama1} K. Washiyama and D. Lacroix, Phys. Rev. C {\bf 78}, 024610 (2008).

\bibitem{Washiyama2} K. Washiyama, D. Lacroix, and S. Ayik, Phys. Rev. C {\bf 79}, 024609 (2009).

\bibitem{Feldmeier} H. Feldmeier, Rep. Prog. Phys. {\bf 50}, 915 (1987).

\bibitem{kim97}
K.-H. Kim, T. Otsuka, and P. Bonche, J. Phys. G {\bf 23}, 1267 (1997).

\bibitem{Randrup2} J. Randrup and W. J. Swiatecki, Ann. Phys. (N.Y.) {\bf 125}, 193 (1980);
Nucl. Phys. {\bf A429}, 105 (1984).

\bibitem{Randrup82} J. Randrup, Nucl. Phys. {\bf A383}, 468 (1982).

\bibitem{denisov02}
V. Y. Denisov and W. N\"orenberg, Eur. Phys. J. A {\bf 15} (2002).

\bibitem{Fre84} H. Freiesleben and J. V. Kratz, Phys. Rep. {\bf 106}, 1 (1984).

\bibitem{Ada94} G. G. Adamian, A. K. Nasirov, N. V. Antonenko, and R. V. Jolos, Phys. Part. Nucl. {\bf 25}, 583 (1994).

\end{thebibliography}
\end{document}